\documentclass[11pt]{article}
\usepackage{authblk}

\usepackage[text={7.5in,9.5in},centering]{geometry}
\usepackage{amsmath}
\usepackage{graphicx}
\usepackage{float}
\usepackage{color}
\usepackage{colortbl}
\usepackage{verbatim}
\usepackage{array}

\providecommand{\e}[1]{\ensuremath{\times 10^{#1}}}

\DeclareMathAlphabet{\mathsfsl}{OT1}{cmr}{bx}{it}

\begin{document}


\title{A computational study of chemically heterogeneous particles: \\
patchy vs. uniform particles in shear flow}

\author[*]{Marina Bendersky}
\author[**]{Maria M. Santore}
\author[*]{Jeffrey M. Davis}
\affil[*]{Department of Chemical Engineering, University of Massachusetts Amherst}
\affil[**]{Department of Polymer Science and Engineering, University of Massachusetts Amherst}

\renewcommand\Authands{ and }




\maketitle

\begin{abstract}

The adhesion of flowing particles and biological cells over fixed collecting surfaces is vitally important in diverse situations and potentially controlled by small-scale surface heterogeneity on the particle.  
Differences in the behavior of patchy particles (flowing over uniform collectors) relative to the reverse case of uniform particles (flowing over patchy collectors) are quantified. 
Because a particle rotates more slowly than it translates in the shear field near a collecting surface, the effective interaction time of a patch on a particle is larger than that of a patch on the collector, suggesting distinct
particle capture tendencies in each case. 
This paper presents a new computational approach to simulate the near-surface motion (rotation and translation) of particles having nanoscale surface heterogeneities flowing over uniform collectors.
Small amounts of ~10 nm cationic patches randomly distributed on a net-negative particle surface produced spatially varying DLVO interactions that were computed via the Grid Surface Integration (GSI) technique and then combined with hydrodynamic forces in a mobility tensor formulation.
Statistical analysis of simulated trajectories revealed fewer extrema in the fluctuating particle-collector separation of heterogeneous particles, compared with the reverse system geometry of uniform particles flowing past a heterogeneous fixed surface.  
Additionally, the patchy particles were captured to a lesser extent on uniform surfaces compared with the case of uniform particles flowing above patchy collectors. Such behavior was dependent on ionic strength, with the greatest differences obtained near a Debye length of $\kappa^{-1} = 4$ nm for the $2a = 500$ nm simulated particles.

\end{abstract}


\section{Introduction}

The adhesive capture of flowing particles on fixed surfaces is a critical step in the transport of bacteria in groundwater \cite{BacteriaAdhesion:2007, bacteriaTransportNelsonGinn:2005}, in the initial stages of innate immune response \cite{AdheDynSimsHammer:1992}, and in the manipulation of particles for applications ranging from mining to microfluidic sensors \cite{LafleurMicroFluidicSensors2012}.    
In such processes, flowing particles (or cells) are typically understood in terms of uniform properties \cite{Adamczyk:2001, DuffadarDavis:2007, DuffadarDavisFriction:2008}: surface charge, hydrophobicity/hydrophilicity balance, and van der Waals interactions. 
Spatial distribution of surface functionality, ``surface heterogeneity", 
on these naturally-occurring and technologically relevant particles is consistently neglected, at length scales that vary from a few nanometers to the particle radius \cite{SongJohnEli:1994, Riehle:2008}.

Patterned colloids such as Janus particles \cite{YoshidaLahann:2008, AlexeevUspalBalazs:2008} and colloidal molecules \cite{GlotzerSolomon:2007} have attracted intense interest for their potential to assemble into ordered periodic structures, not only for photonic applications \cite{PawarKretzschmarPatchyReview:2010, ShevchenkoEtAl:2008}, but also, more fundamentally, to mimic the phase transitions of molecules \cite{RussoEtAl:2009, TavaresEtAlMolecPhys:2009, TavaresEtAl:2009}.  
The ability of engineered chemical and topographical patterns on these particles to direct their interactions suggest \cite{ZhangGlotzer:2004, GlotzerSolomon:2007} that the randomly distributed heterogeneities on naturally-occurring particles will influence their adhesion as well, even though the length scales of such particle features fall well below the colloidal dimensions \cite{SantoreKozlova:2006, SantoreKozlova:2007}.

The literature of the past two decades contain numerous efforts that model the adhesion of flowing particles on surfaces regularly-patterned with features that approach the lengthscales of the flowing particles \cite{Hydrobump:2003}. 
The adhesion of flowing micron-sized particles on nano-scale heterogeneous collectors, however, has only been examined recently \cite{Adamczyk:2001, DuffadarDavis:2007, DuffadarDavisFriction:2008, KempsBhattacharjee:2005}.
Surface heterogeneity, defined as a random arrangement of surface features, 
was shown to influence particle adhesion significantly, and it was demonstrated, for instance, that randomly-positioned clusters of nanoscale cationic patches on the collector can drive colloidal particle deposition onto net negative collecting surfaces \cite{SantoreKozlova:2006}.  
Even when each nanoscale cluster or ``patch" of positive charge, positioned on the negative background surface, is individually too weakly attractive towards uniform flowing anionic particles, the concerted action of many of such patches can capture particles. 
Modeling methods that describe particle-collector interactions have evolved considerably over the past decade, from analytical \cite{SureshWalz:1996} and numerical \cite{BhattacharjeeKoElimelech:1998, SunWalz:2001} computations of systems characterized by surface roughness, to the development of the surface-element integration (SEI) \cite{BhattacharjeeElimelechSEI:1997} and the more recent grid-surface integration (GSI) \cite{BenderskyDavis:2011, DuffadarDavis:2007, DuffadarDavisFriction:2008, DuffadarDavisSantore:2009} techniques.
While the SEI technique involves the discretization of only one interacting surface, in the now-accepted \cite{MaPazminoJohnson:2011, BradfordTorkzaban:2013, BradfordTorkzabanAdhParam:2013} GSI technique
surface forces are integrated over differential elements on two opposing objects of arbitrary shapes.

In many recent computational studies, patchy particles are modeled as units composed of distinct ``atoms.'' Specific attributes are assigned to each atom, depending on whether it belongs to the `patch' or `core' surface area of the particle. Zhang and Glotzer\cite{ZhangGlotzer:2004} performed Brownian dynamic simulations to model the self assembly of particles patterned with patches positioned at specific locations. 
Polyelectrolytes of varying shapes are also shown by molecular dynamic computations to assemble into charged, patchy colloids \cite{LikosEtAl:2008}.

While theoretical work on patchy particle interactions at the molecular scale abound, only a handful of studies describe colloidal interactions of heterogeneous particles flowing near a planar wall and the effects of such heterogeneities on particle deposition. 
In contrast to the studies on heterogeneous fixed collectors, the system geometry comprising heterogeneous particles presents the additional complexity that, in shear flow, the particle rotates, exposing different areal elements to the collecting surface as it flows past. This rotation and turnover in the dominant surface region poses a computational challenge in the calculation of time-dependent colloidal interactions.  
Only a limited number of studies have addressed computationally related problems: Sphere-plate and sphere-sphere DLVO interactions for spheres patterned with roughness (topographical heterogeneity) were computed numerically \cite{BhattacharjeeKoElimelech:1998, SunWalz:2001} and analytically \cite{SureshWalz:1996} and found to agree with experimental measurements \cite{SureshWalz:1997}. More recently, Chatterjee et al. focused on the deposition of micro-particles onto Janus and patchy spherical collectors \cite{ChatterjeeMitraBhattacharjee:2011}. Chemical heterogeneity was modeled by patterning the spherical collector with adhesion-favorable and adhesion-unfavorable surface properties assigned to the Janus collectors or with relatively wide alternating stripes (much wider than the particle radius) that covered the spherical surface. Deposition of the colloidal particles onto spherical heterogeneous collectors was analyzed as a function of collector orientation, collector patterning, and colloidal particle velocity.

Similarly, the adhesion of leukocytes on the vascular endothelium was studied computationally by Korn and Schwarz \cite{KornSchwarz:2008}, who modeled flowing leukocytes that were captured by the binding of randomly-positioned immobilized receptors. The adhesive receptors on the leukocytes, modeled as hard spheres, were similar to the randomly distributed nano-patches on flowing micro-particles. The leukocyte capture, however, is fundamentally different from the heterogeneous particle adhesion, since the receptors underwent chemical binding according to a reaction probability when in close distance to a surface ligand. Interesting motions of leukocytes, that ranged from free flow to firm adhesion, were presented as a function of the attachment and detachment rates of the cells to the wall in adhesion regime diagrams.
The treatment lacked the spatially varying longer range colloidal forces of interest in the current study, but was, nonetheless, a break-through in its attention to patterned particle rotation in flow. As a result of this advance, the computations predicted interesting motion signatures such as particle rolling and skipping.

The present work describes a new computational approach to calculate particle-surface interactions, trajectories and adhesion probabilities of flowing heterogeneous particles on fixed planar uniform collectors.  
The method expands the implementations of the Grid-Surface Integration (GSI) technique by discretizing the particle surface, in order to include on its surface a random distribution of patches. 
DLVO (electrostatic and van der Waals) interactions are computed within the GSI technique on the basis of parallel flat plate expressions \cite{Hamaker:1937, Hogg:1966}, and flow-driven motion is obtained from the mobility matrix of the hydrodynamics problem \cite{DuffadarDavis:2007}. Two heterogeneous particle-collector system geometries are, moreover, compared in detail. 
While one of such systems comprises electrostatically heterogeneous particles flowing past a uniform anionic collector, the alternative system consists of uniform anionic particles flowing past an electrostatically heterogeneous collector, such that surface compositions in each system are reversed.  
The two configurations differ in their number of particle trajectory fluctuations and their collection probability due to the particles' rotational motion, given by the rotational velocity times the particle radius, that is slower than its translational motion, determined by the translational velocity. 
It is found that patchy particles present less fluctuating trajectories and smaller collection probabilities than those of uniform particles flowing past heterogeneous collectors. 
The findings are presented in terms of statistical parameters of the particle trajectories and adhesion regime diagrams.

\section{Sphere Discretization}
\label{sec:eqspalgthm}

In previous work, the collector surfaces were electrostatically and topographically heterogeneous, while the flowing colloidal particles were smooth, uniformly charged spheres. A discretization of the spherical surface was therefore not needed, and only the heterogeneous collector was partitioned into small areal elements. The accurate modeling of systems of heterogeneous particles, however, does require the discretization of the spherical surfaces into differential elements, each of which can be assigned distinct surface properties. 

\begin{figure}
\centering
\includegraphics[width=0.6\textwidth,natwidth=1054,natheight=1212]{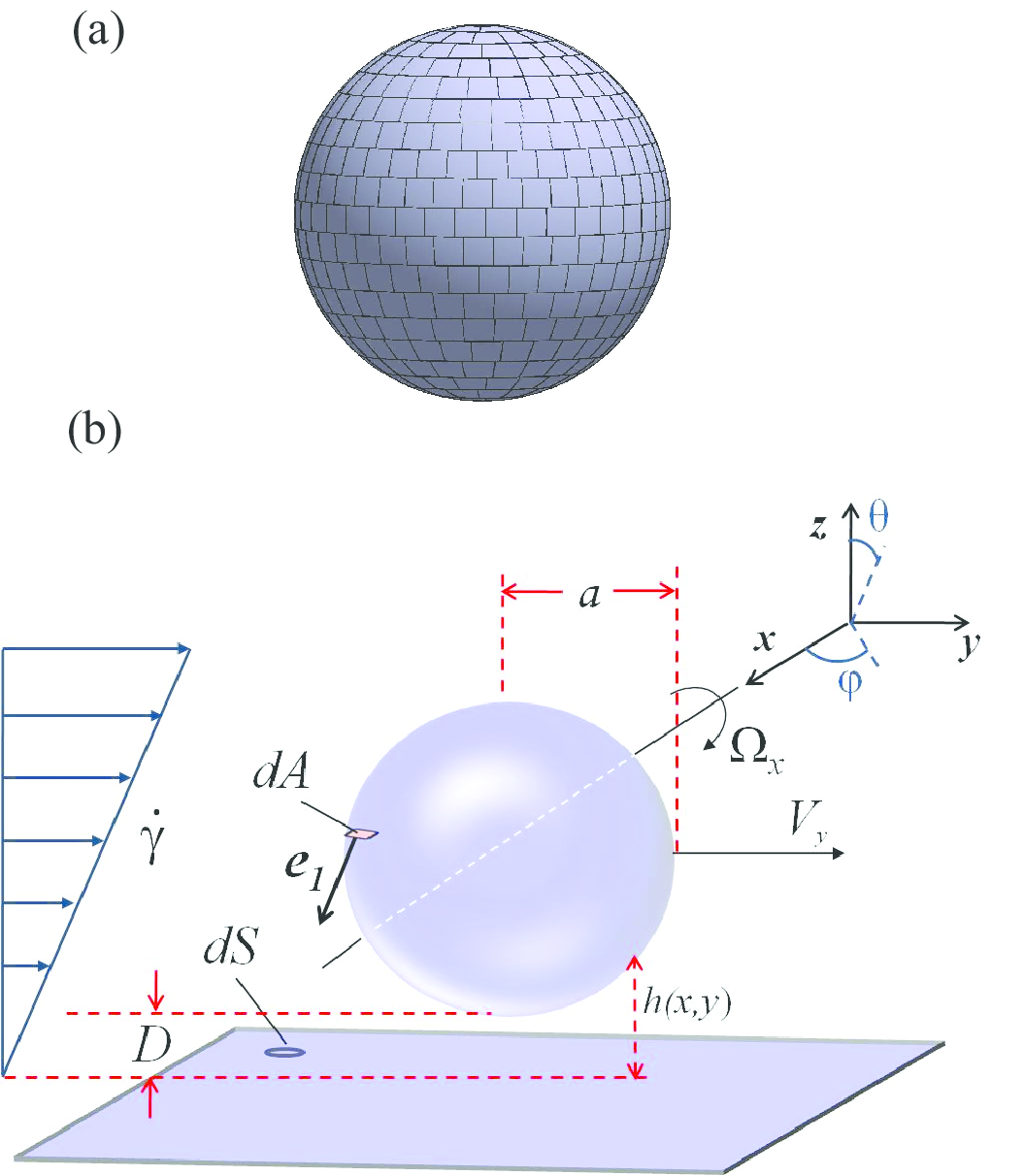}
\caption{(a) Illustration of a sphere discretized into $N_\text{p}$ = 500 equal area regions. (b) Schematic diagram of a uniform particle of radius $a$ interacting with a uniform flat collector. The local and minimum particle-collector separation distances are denoted by $h$ and $D$, respectively. Differential surface elements $dA$ on the particle and $dS$ on the collector are also indicated, as well as the unit vector $\mathbf{e_1}$ that points in the direction of the collector surface element.
The rotational and translational velocities are $\Omega_x$ and $V_y$, respectively, and the flow shear rate is $\dot{\gamma}$.}
\label{fig:Fig1}
\end{figure}

In the results presented in this work, spherical surfaces are discretized into regions of equal area using Leopardi's \cite{Leopardi:2006} recursive zonal EQual area Sphere Partitioning (EQSP) algorithm.  A schematic diagram of a sphere discretized into $N_\text{p}$ = 500 regions (elements) is presented in Fig. \ref{fig:Fig1}(a).

The accuracy of the EQSP algorithm is first verified by discretizing a $2a = 1 \,\mu m$ diameter sphere into $N_\text{p} = 25963$ elements, such that the individual element area is d$A_i \simeq$ 121 nm$^2$, corresponding to an 11 nm $\times$ 11 nm patch for consistency with previous studies \cite{BenderskyDavis:2011} motivated by experiments \cite{SantoreKozlova:2006, SantoreKozlova:2007}. For each surface element on the sphere, its projected area on an horizontal plane is computed from
\begin{equation}
dS_i = |\mathbf{n} \cdot \mathbf{e}_\perp|\,dA_i\,,
\end{equation}
where $dA_i$ is the area of the sphere element $i$, $dS_i$ is its projected area on the plane, $\mathbf{n}$ is the vector normal to the sphere element and $\mathbf{e}_\perp$ denotes the direction normal to the projection plane. The sum of the elemental projected areas $\sum dS_i$ is within 0.0038\% of the exact value of 2$S = 2\pi\,a^2$ (for two hemispheres).

\section{The GSI technique}

In the GSI technique, the total force or energy of interaction is obtained from a pairwise summation of interactions between differential areal elements on the collector and particle surfaces \cite{DuffadarDavis:2007, DuffadarDavisFriction:2008}:
\begin{equation}
F = \sum_{\mathrm{particle}} \sum_{\mathrm{wall}}\; P(h)\; \mathsfsl{e}_{1} \cdot\mathsfsl{e}_z \;  dS,
\label{eqn:GSIdoublesum}
\end{equation}
where $P(h)$ is the force or energy of interaction per unit area between a particle's areal element ($dA$) and a corresponding element on the substrate ($dS$). The unit vector $\mathsfsl{e}_z$ specifies the direction normal to the surface, and the unit vector $\mathsfsl{e}_1$ indicates the direction between areal elements on the particle and the substrate (see Fig.\ref{fig:Fig1}(b)). The summations in Eq. (\ref{eqn:GSIdoublesum}) are performed over all of the discretized elements on the particle and collector.
The colloidal energy of interaction per unit area ($P(h) = U^A(h)$) between each pair of areal elements is obtained by summing van der Waals (vdW) and electrostatic double layer (EDL) components, each of them computed from analytical expressions \cite{Hamaker:1937, Hogg:1966}  derived for parallel plates.
The corresponding forces per unit area ($P(h) = F^A(h)$) are obtained by differentiation of the energies with respect to the separation distance and included in the mobility matrix formulation to yield particle trajectories \cite{DuffadarDavis:2007, DuffadarDavisFriction:2008, BenderskyDavis:2011}. 

Within the GSI technique, each areal element on each interacting surface can be assigned distinct properties, such that GSI computations can be implemented for particles and collectors of arbitrary shape, topography and chemical properties.

\section{Uniform particle-collector systems}
\label{sec:homosys}

A schematic diagram of a uniform particle-collector system, in which all interacting surface elements are assigned the same properties, is presented in Fig. \ref{fig:Fig1}(b). 
Negatively charged, flowing particles of radius $a$ interact with the flat uniform collector substrate, which is also negatively charged.
The local and minimum particle-collector separation distances are denoted by $h$ and $D$, respectively.
Due to the linear shear flow with shear rate $\dot{\gamma}$, the particle translates in the $y$-direction with velocity $V_y$ and rotates around an axis parallel to the collector surface with a rotational velocity $\Omega_x$.
The origin of the Cartesian coordinate system is the left edge of the collector.
For interactions of $2a = 1 \mu$m diameter particles, the collector length is $L = 30 \mu$m, while shorter collectors of length $L = 20 \mu$m were simulated for interactions of $2a = 500$ nm diameter particles.
A spherical coordinate system with origin in the sphere's center, is also defined.
The inclination angle is $\theta$, $0 \le \theta \le \pi$, and the azimuthal angle is $\phi$, $0 \le \phi \le 2\pi$. The radial direction is normal to the particle surface.

Electrostatic and van der Waals interactions for the uniform system illustrated in Fig. \ref{fig:Fig1}(b) are obtained by implementing the GSI technique, and including the discretization of either one or both interacting surfaces. For the case of a uniform spherical particle, all the EQSP-generated differential elements are assigned the same surface properties, which also equal those of the uniformly patterned collector.

\begin{figure}
\centering
\includegraphics{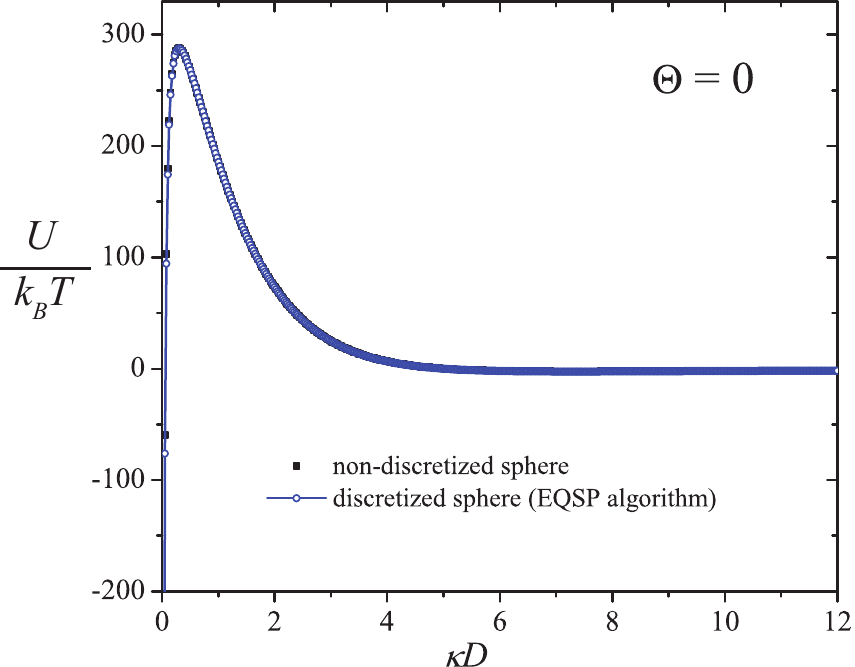}
\caption{Energy-distance profile for a uniform particle interacting with a uniform collector. The particle's surface is either discretized into equal area elements (with the EQSP algorithm) or treated as a non-discretized uniform surface. In both cases, DLVO interactions are computed with the GSI technique. The simulation parameters are: $2a$ = 1 $\mu$m, $\dot{\gamma}$ = 25 s$^{-1}$, $\kappa^{-1}$ = 4 nm, $A_\text{H}$ = 5\e{-21}, $\Psi_\text{collector} = \Psi_\text{sphere}$ = -25 mV.}
\label{fig:Fig2}
\end{figure}

In Fig. \ref{fig:Fig2}, the energy profile of a smooth and uniformly charged 2$a = 1\,\mu m$ diameter particle interacting with a flat surface is shown. Colloidal interactions are characterized by the Hamaker constant $A_\text{H}$ = 5\e{-21} J, which is fixed throughout this study, and the
inverse Debye screening length $\kappa^{-1}$ = 4 nm. The surface loading $\Theta$, defined as the area of the collector 
containing attractive (cationic) functionality relative to the total collector area, is equal to zero for both interacting surfaces, which carry an electrostatic potential $\Psi = -25$ mV. The energy profile is thus defined by the repulsive electrostatic double layer (EDL) interactions between the uniformly and equally charged particle and collector surfaces.

The interactions presented in Fig. \ref{fig:Fig2} are computed by implementing the GSI technique and either including (solid line) or not (dotted line) the EQSP algorithm for the modeling of the uniformly charged, smooth sphere.
As in Sec. \ref{sec:eqspalgthm}, the uniform $2a = 1 \mu$m diameter particle is discretized into $N_\text{p} = 25963$ areal elements, such that the area of each element is d$A_i \simeq$ 121 nm$^2$. In the case for which the spherical surface is not discretized into differential elements, the collector surface is partitioned instead. The collector grid consists of 91 square elements that represent a length of $2a = 1\,\mu$m, such that the length of each square element is $\approx 11$ nm. 
The spherical and planar discretization schemes are specifically chosen so as to define equal-area elements, in this case of d$A_i \simeq$ 121 nm$^2$. In heterogeneous systems, the size of one areal element can sometimes equal the surface feature size \cite{BenderskyDavis:2011} though it should be smaller if the surface chemical patterns become relatively large. Discretization schemes, therefore, vary with particle and surface features dimensions.

It is seen in Fig. \ref{fig:Fig2} that both computational techniques yield results that are in perfect agreement, thus validating the implementation of the \emph{GSI-EQSP} technique.

\section{Random distribution of cationic patches on the sphere.}
\label{sec:hetsysrand}

\begin{figure}
\centering
\includegraphics{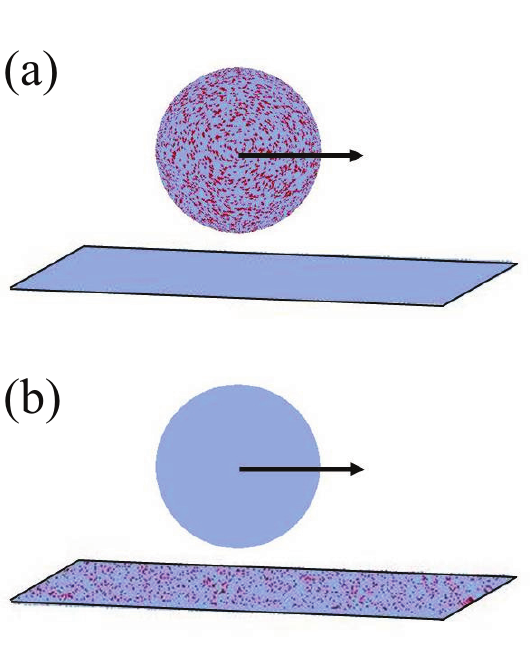}
\caption{Schematic diagrams of particle-collector systems with one heterogeneous surface. (a) Patchy sphere-uniform collector. (b) Uniform sphere-patchy collector.}
\label{fig:Fig3}
\end{figure}

Particles and collectors with randomly positioned attractive cationic patches can be constructed by numbering the discretized sphere/collector elements and then choosing the location of each patch sequentially by mapping a uniform deviate. 
In this section, interactions between patchy particles and uniformly charged, flat collectors are studied for different particle sizes, surface loadings and Debye lengths.
Detailed comparisons to interactions in uniform particle-patchy collector systems are also presented.
The patchy particle and patchy collector systems are schematically depicted in Fig.\ref{fig:Fig3}.
Patches on either heterogeneous surface are assigned a potential of $\Psi_\text{het} = 50$ mV, while the other regions of the heterogeneous surface and the homogeneous surface bear a uniform electrostatic potential of $\Psi_\text{uni} = -25$ mV. 
The interactions between a patch and the uniform surface are therefore attractive, while the interactions between other regions are repulsive.
The individual patch area equals that of the surface element (A$_\text{patch} \simeq 121$ nm$^{2}$), such that the patches model 11 nm squares.
Due to the Poisson distribution used to locate patches on both heterogeneous surfaces (planar or spherical), there are no regions on either heterogeneous surface where patches are preferably assigned.

\begin{figure*}
\centering
\includegraphics[width=6.5in]{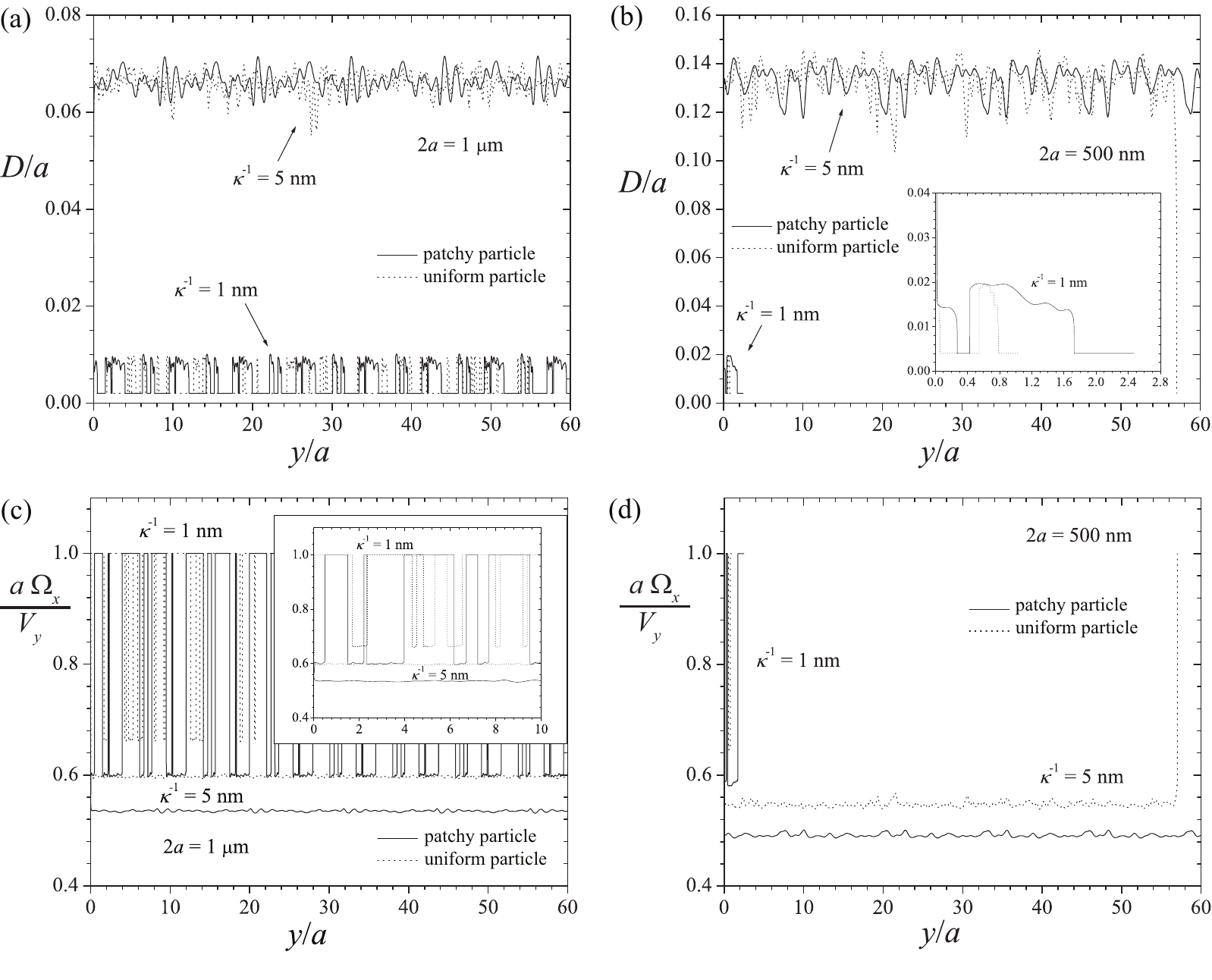}
\caption{Trajectories and velocity ratios for patchy and uniform particles of sizes $2a = 0.5, 1 \mu$m interacting at Debye lengths $\kappa^{-1} = 1, 5$ nm, for a fixed surface loading $\Theta = 0.17$. Rolling friction is computed with a friction coefficient of $\mu_\text{R} = 1.3 \e{-4}$ \cite{DuffadarDavisFriction:2008}. (a)(b) Particle trajectories of patchy and uniform particles, for particle sizes of $2a = 1 \mu$m (a) and $2a = 500$ nm (b). (c)(d) Angular to translational velocity ratios of patchy and uniform particles, for particle sizes of $2a = 1 \mu$m (c) and $2a = 500$ nm (d).
For clarity, all solid lines represent the patchy particle system, while all dotted lines indicate results obtained for the uniform particle (patchy collector) system.}
\label{fig:Fig4}
\end{figure*}

For a fixed area fraction of cationic patches of $\Theta = 0.17$, trajectories of patchy particles flowing over uniformly charged collectors and those of uniformly charged particles flowing over patchy collectors
are presented in Fig. \ref{fig:Fig4}(a)-(b) for two particle sizes and Debye lengths.
For each case, the ratio of the particle's rotational to translational velocities are plotted as a function of the horizontal displacement in Fig. \ref{fig:Fig4}(c)-(d). For clarity, Fig.\ref{fig:Fig4}(c) and the inset in such figure are presented separately in the Supporting Information.

In particle trajectory and velocity calculations presented in Fig. \ref{fig:Fig4}, the particle is assumed to contact the collector at an arbitrarily small separation distance $D = \delta = 1$ nm that models the surface roughness \cite{DuffadarDavis:2007, DuffadarDavisFriction:2008}.
When the separation distance $D < \delta$, the particle 
either rolls in contact with the collector, or, arrests due to friction forces.
The condition for particle arrest is found from a force balance in the direction of flow and reads \cite{DuffadarDavisFriction:2008}
\begin{equation}
F_y + \frac{T_x}{a} - F_{rf} < 0
\end{equation}
where $F_y$ and $T_x$ are the shear-induced force and torque \cite{DuffadarDavis:2007, DuffadarDavisFriction:2008}, respectively, and $F_{rf} = \mu_R F_z^N$ is the rolling resistance, determined by a coefficient of rolling friction $\mu_R$ and the normal force $F_z^N = -F_{DLVO}$. The particle therefore arrests on the collector once the DLVO forces, multiplied by an appropriate friction coefficient, are sufficiently attractive to overcome the shear-induced effects, or, when
\begin{equation}
F_y + \frac{T_x}{a} + \mu_R F_{DLVO} < 0\,.
\end{equation}
The rolling friction is computed with a friction coefficient of $\mu_{R} = 1.3 \e{-4}$ \cite{DuffadarDavisFriction:2008}.

In Fig. \ref{fig:Fig4}(a), the upper lines represent trajectories of $2a = 1 \mu$m diameter patchy and uniform spheres translating in shear flow above uniform and patchy collectors respectively, for a Debye length of $\kappa^{-1} = 5$ nm.
Due to the large Debye length, both particles translate at a relatively large separation distance from the collector, and the particles do not contact or deposit on the collector.
The average separation (mean $\pm$ standard deviation) for the patchy particle is $D/a = 6.64\e{-2} \pm\, 2.08\e{-3}$, 
while that of the uniform particle is $D/a = 6.60\e{-2} \pm\,2.13\e{-3}$,  
in agreement with the secondary minimum in the energy profile 
$D_\text{min}/a = 6.62\e{-2}$, computed with the GSI$_\text{UA}$ technique \cite{BenderskyDavis:2011}.
Although the mean separation distance and its standard deviation are consistent for the patchy and uniform particles, collection probabilities are found to be different in patchy particle systems than in patchy collector systems.
This behavior is examined in more detail in Secs. \ref{sec:CollEff} and \ref{sec:ResTimes}.
In Fig. \ref{fig:Fig4}(c), the ratio of the rotational and translational velocities for a Debye length $\kappa^{-1} = 5$ nm are shown by the lower solid and dotted lines, which indicate results for the patchy and uniform spheres, respectively. In both cases, the ratios 
fluctuate around the same values, that are less than unity when the particle is not in contact with the collector, confirming that the particles translate faster than they rotate.

The lower solid and dotted lines in Fig. \ref{fig:Fig4}(a) show trajectories of $2a = 1 \mu$m diameter patchy and uniform particles, respectively, for a Debye length $\kappa^{-1} = 1$ nm. It is seen that both particles frequently contact the collector surface (at an arbitrarily small distance $\delta = 1$ nm taken as representative of surface roughness \cite{DuffadarDavisFriction:2008}) to yield a trajectory that is characterized by alternating periods of rolling motion and of free flow in close proximity to the collector. Both particles maintain a separation distance $D < 5$ nm as they flow above the entire simulated collector. Velocity ratios as a function of the horizontal displacement are shown by the upper solid and dotted lines (for patchy and uniform particles respectively) in Fig. \ref{fig:Fig4}(c).
The rolling periods for both particles are identified by velocity ratios that are equal to unity, as it is assumed that the particle does not slip when it contacts the surface \cite{CoxBrennerI:1967}, while the translation between those periods corresponds to velocity ratios that fall below unity.

Fig. \ref{fig:Fig4}(b) presents trajectories of $2a = 500$ nm particles at Debye lengths $\kappa^{-1} = 1, 5$ nm. The patchy spheres are discretized into $N_\text{p} = 6489$ elements, such that the surface area of each element remains d$A_i \simeq$ 121 nm$^2$, and the patchy collectors are modeled with a grid consisting of 45 square elements that represent a length of $2a = 500$ nm, such that the length of each square element is $\approx 11$ nm. At a large Debye length of $\kappa^{-1} = 5$ nm, the patchy particle (upper solid line) translates at a separation $D/a$ of $(0.134 \pm 5.8\e{-3})$, while that of the uniform particle (upper dotted line) is $(0.133 \pm 6.72\e{-3})$, in good agreement with the secondary minimum in the energy profile of D$_\text{min}/a$ = 0.134, computed with the GSI$_\text{UA}$ technique \cite{BenderskyDavis:2011}. The amount of spatial fluctuations of the uniform particle trajectory is larger, as expected from the larger 
standard deviation of the separation distance. More importantly, the uniform particle (upper dotted line) adheres on the patchy collector, while the patchy particle (solid upper line) flows above the entire uniform collector, without adhering. The respective velocity ratios, presented in the lower solid and dotted lines in Fig. \ref{fig:Fig4}(d) are smaller than 1. Both particles translate fast, with respect to their rotational motion, and, for the uniform particle, the ratio increases to 1 when the particle is arrested on the collector due to a rolling resistance in the direction of flow. Such rolling resistance retards the particle's motion as a consequence of elastic deformations of the surfaces in contact \cite{DuffadarDavisFriction:2008}.

For a Debye length of $\kappa^{-1} = 1$ nm, trajectories of patchy and uniform $2a = 500$ nm diameter particles are shown by the lower solid and dotted lines in Fig. \ref{fig:Fig4}(b), and enlarged for clarity in the inset of the same figure. Both particle trajectories exhibit rolling periods that alternate with free flow segments in which the particles translate in close proximity to the collector.  Due to friction forces, both particles ultimately adhere on the collector when the DLVO attraction becomes sufficiently large. The velocity profiles, denoted by the upper solid and dotted lines in Fig. \ref{fig:Fig4}(d), reach a value of unity, thus indicating rolling periods, but fluctuate around values lower than unity in the trajectory segments in which the particles loose contact with the collector.

\subsection{Collection Probability}
\label{sec:CollEff}

In Sec. \ref{sec:hetsysrand}, a few characteristic examples of particle trajectories were computed for different patchy/uniform particle sizes and Debye lengths. The results presented in this section are statistically significant collection probabilities obtained from a large number of computed trajectories.

In the present study, collection probabilities are defined as the ratio of adhered particles (successes) to the total number of simulated particle trajectories (trials),
\begin{equation}
{\eta} = \frac{N_\text{D}}{N_\text{total}}\,
\label{eqn:eff1}
\end{equation}
where $N_\text{D}$ is the number of particles deposited on the collector and $N_\text{total}$ is the total number of particle trajectories simulated. 
The collection probability estimate given by Eq. (\ref{eqn:eff1}) is the ratio of successful adhesion attempts to the total number of attempts and resembles adhesion probabilities \cite{KempsBhattacharjee:2009} or the available surface function (ASF) defined as the normalized particle adsorption probability in the limit N$_\text{attempts}\,\rightarrow\,\infty$ \cite{Adamczyk:2002}.
It should be noted that, in the present study, particle trajectories were computed sequentially for collectors with no adhered particles and particle adhesion is thus determined by particle translation and rotation in shear flow subject to the computed electrostatic and van der Waals interactions.

\begin{figure}
\centering
\includegraphics{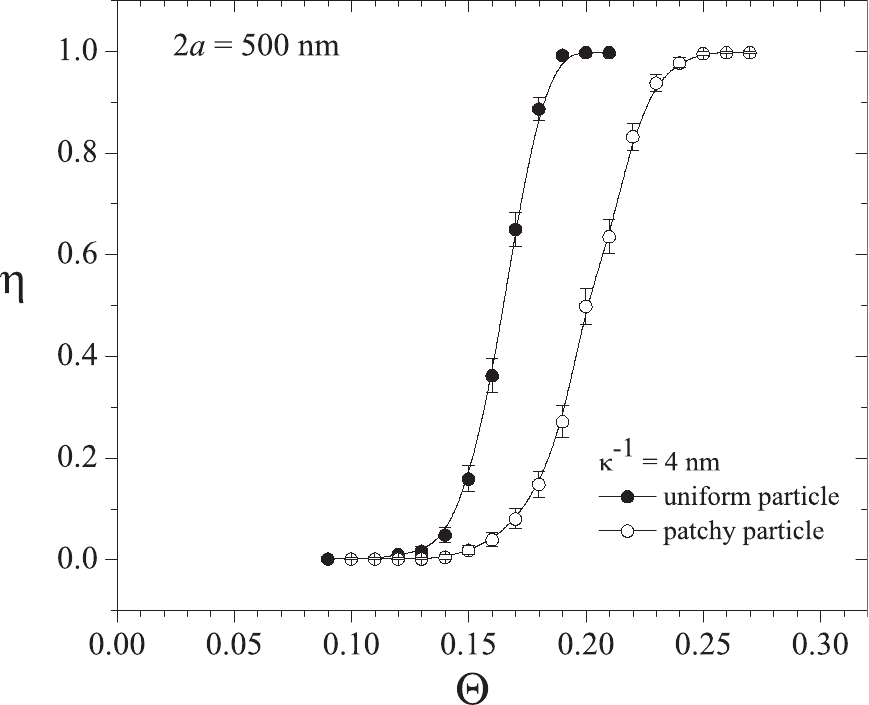}
\caption{Collection probability curves of $2a = 500$ nm diameter, patchy particles adhering to a uniform anionic collector and of uniform anionic particles adhering to a patchy collector. $\Theta$ is the area fraction of positive patches on either patchy surface.}
\label{fig:Fig5}
\end{figure}

In Fig. \ref{fig:Fig5}, collection probability curves are presented for patchy and uniform particles at a Debye length $\kappa^{-1} = 4$ nm.
The total number of particle trajectories simulated for each point on this plot
is $N_\text{total} = 1000$.
Wilson score intervals \cite{Wilson:1927} are constructed to provide more statistically relevant predictions and the error bars in Fig. \ref{fig:Fig5} correspond to the 95\% confidence interval.
Each of the trajectories computed for patchy particles was performed for a different heterogeneous particle in which the patches are randomly distributed on the spherical surface. Alternatively, uniformly charged particles were simulated to flow over distinct heterogeneous collectors, each of which was patterned with patches located at randomly chosen locations. Once the sphere or planar surface was discretized into elements with equal area, the algorithm used to choose the location of a patch was identical.

Unlike deposition rate curves, the collection probability curves presented in Fig. \ref{fig:Fig5} do not contain any rate information.
In similar computational studies performed on the basis of the Random Sequential Adsorption (RSA) model, the particle adsorption kinetics is characterized by plots of the absorbed particle coverage as a function of a dimensionless  adsorption time that depends on the number of simulated adhesion attempts \cite{AdamczykRSA:1998, Adamczyk:2002}, which is most relevant when a significant fraction of the surface is covered by adhered particles. 
Particle deposition modeled with RSA simulations does not include the effects of DLVO interactions, but instead depend on a series of attempts in which locations on the heterogeneous surface are randomly chosen. 
A particle will be considered to be irreversibly adhered at a specific location if an unoccupied adhesion site is present at such location, and if the additional particle would not overlap with previously adhered ones.
Experimental work \cite{KalasinSantore:2008, SantoreRolling:2010, SantoreKozlova:2006, SantoreKozlova:2007} involving heterogeneous collectors with nanoscale patches has been focused on adhesion thresholds and initial deposition rates (when the fraction of the collector covered by adhered particles is negligible). In the present study,  
each particle was thus sampled one at a time and each adhesion attempt is represented by a different (uniform or patchy) particle which adheres, or not, on the (patchy or uniform) collector, which is not occupied by previously adhered particles.
Particle adhesion is then determined by DLVO particle-collector interactions, as the particle translates and rotates in shear flow.

The collection probability curves presented in Fig. \ref{fig:Fig5} provide an estimate of the particle adhesion threshold, defined as the smallest surface loading for which particles adhere on the collector ($\Theta_c = \text{min}(\Theta): {\eta} >0$). 
Interestingly, for the chosen set of parameters, the adhesion threshold for the patchy sphere on a uniform collector ($\Theta_c = 0.14$) is larger than that of the uniform sphere on a patchy collector ($\Theta_c = 0.12$). The higher threshold for the patchy sphere is attributed to its decreased tendency to contact and adhere on the uniform collector, with respect to the uniform sphere flowing over the patchy collector.
The patchy particle is less prone to adhere on the uniform collector due to its larger patch interaction time, as described in detail in Sec. \ref{sec:ResTimes}.

\subsection{Number of local extrema in particle-collector separation}

Trajectories for the patchy particle and patchy collector systems differ in the extent of fluctuations in the particle-collector separation.
Such differences can be quantified by computing the number of local extrema in the trajectories as a function of the surface loading $\Theta$, for each particle type (patchy or uniform).
Due to the inherent random nature of the heterogeneity, the number of extrema for each particle type at each value of $\Theta$ is obtained as an average performed over many trajectories.

Since not all particle trajectories denote particle adhesion, the average density of extrema for a fixed $\Theta$ is obtained as the linear combination
\begin{equation}
<\overline{N_\text{ext}}> = p \, <\overline{N_\text{ext}}>_{a} + (1-p) \, <\overline{N_\text{ext}}>_{na}
\label{eqn:NextAvg}
\end{equation}
where
the subscripts $a$ and $na$ denote `adhesion' and `no adhesion' respectively, 
\begin{equation}
p = \frac{N_D}{N_\text{total}}\,,
\end{equation}
$N_D$ is the number of particles deposited on the collector, and
$N_\text{total} \simeq 200$. 
Eq. (\ref{eqn:NextAvg}) is derived in detail in the Supporting Information.
The variance of $<\overline{N_\text{ext}}>$,
\begin{equation}
\sigma^2_{\text{N}_\text{ext}} = p^2 \sigma^2_{\text{N}_\text{ext,\,a}} + (1-p)^2 \sigma^2_{\text{N}_\text{ext,\,na}}\,,
\end{equation} 
is obtained from an error propagation expression derived for Eq. (\ref{eqn:NextAvg}), assuming that, for the specific case of the trajectories considered in these results, $p$ is a precisely known constant.

\begin{figure}
\centering
\includegraphics{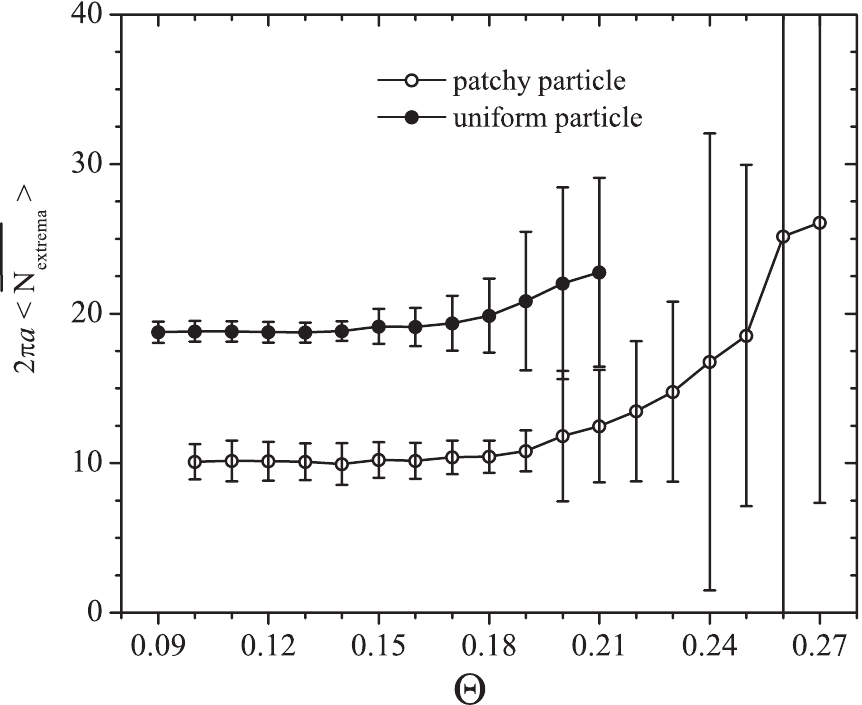}
\caption{Weighted number of extrema as a function of $\Theta$ for patchy and uniform $2a = 500$ nm particles, averaged over $N_\text{tot} \simeq 200$ trajectories.}
\label{fig:Fig6}
\end{figure}

In Fig. \ref{fig:Fig6}, the number of extrema is plotted as a function of the surface loading $\Theta$ for $\kappa^{-1} = 4$ nm.
It is seen that for both patchy and uniform particles the average number of local extrema remains relatively constant for low values of the surface loading of patches $\Theta$, but increases as the surface loading is further increased.
As described in Sec. \ref{sec:hetsysrand}, interactions with 
patches are defined as attractive. For increasing surface loading of patches, thus, the increasingly stronger attractive interactions decrease the particle-collector separation distance, resulting in a larger number of spatial fluctuations (number of extrema) in the particle trajectories.
The average number of extrema starts increasing at $\Theta \simeq 0.17$ and at $\Theta \simeq 0.19$ for the uniform and patchy particles, respectively.
The standard deviation in the number of extrema increases as well, and more significantly than the number of extrema itself, with increasing surface loading $\Theta$. As the number of adhering particles increases,
the existence of different adhesion sites on each heterogeneous collector multiplies the number of different trajectories that denote particle adhesion, 
thus increasing the standard deviation in $<\overline{N_\text{ext}}>$ at larger values of $\Theta$. As shown in Fig. \ref{fig:Fig6}, the standard deviation increases dramatically when the adhesion probability approaches unity.

The total number of extrema for the patchy particle is smaller ($N_\text{tot,p} \simeq \,N_\text{tot,u}/2$) than that of the uniform particle. 
As described in detail in Sec. \ref{sec:ResTimes}, the larger patch interaction time of the patchy particle partially precludes interactions with other heterogeneity regions, which gives rise to smoother trajectories, or, equivalently, a smaller number of extrema.
For each particle type (patchy and uniform), the maximum and minimum number of extrema are also computed and found to be essentially equal, $N_\text{max,\,p\,(u)} = N_\text{min,\,p\,(u)} = N_\text{total,\,p\,(u)}/2$, as expected for a fluctuating trajectory line.

\subsection{Patch interaction times}
\label{sec:ResTimes}

Differences in the trajectories and adhesive behaviors of patchy and uniform particles can be interpreted in terms of an effective time period over which one patch influences the electrostatic double layer interaction between a particle and the collector.

For simplicity, the patch interaction time is defined as the time a sphere surface element interacts within the zone of influence (ZOI) on the collector. The radius of the ZOI is defined approximately by the overlap of the Debye layers \cite{SantoreKozlova:2006, SantoreKozlova:2007} and given by $R_\text{ZOI} = 2\sqrt{\kappa^{-1}a}$.
The newly defined patch interaction time is, in fact, the \emph{maximum} patch interaction time, since the particle could adhere on the collector by translating a distance that is smaller than the $R_\text{ZOI}$. 
For a uniform particle translating above a patchy collector, the patch interaction time is the time required for the particle to translate a distance of $2R_\text{ZOI}$.  For a patchy particle, translating above a uniform collector, the patch interaction time is the time required for the particle to rotate through the corresponding subtended angle.

\begin{figure}
\centering
\includegraphics{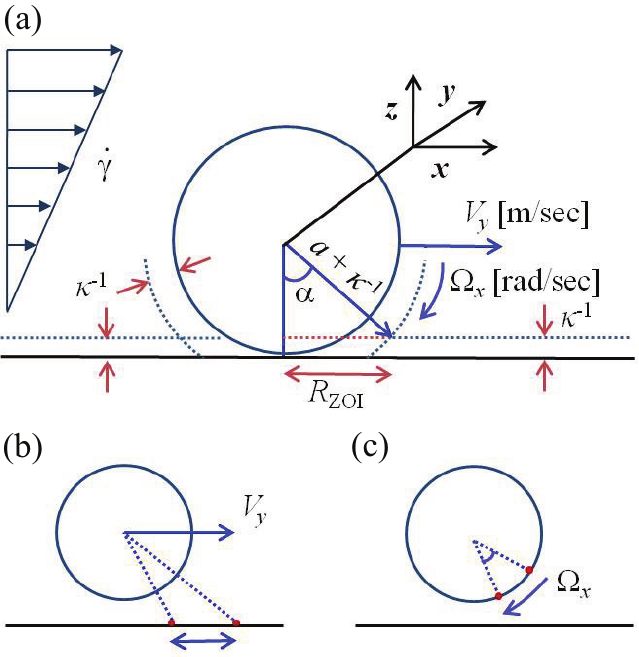}
\caption{(a) Schematic definition of the radius of the Zone of Influence R$_\text{ZOI}$. Debye layers of width $\kappa^{-1}$ around each interacting surface, particle velocities $\Omega_x$ and $V_y$, and the angular displacement $\alpha$ that corresponds to a linear displacement of R$_\text{ZOI}$, are also indicated. (b)(c) Schematic diagrams illustrating the appropriate displacements and velocities that define patch interaction times for patchy collectors and spherical particles. (b) Linear displacement and linear velocity $V_y$ for the case of a patchy collector (uniform particle). (c) Angular displacement and rotational velocity $\Omega_x$ for the case of a patchy particle.}
\label{fig:Fig7}
\end{figure}

A particle in shear flow in close proximity to a planar surface is shown schematically in Fig. \ref{fig:Fig7}(a). The particle translates with velocity $V_y$ and rotates at an angular velocity $\Omega_x$.
For the case in which the patch is located on the collector, the appropriate interaction time depends on the translational velocity $V_y$, as illustrated in Fig. \ref{fig:Fig7}(b), and is defined as
\begin{equation}
\tau_\text{tr} = \frac{2R_\text{ZOI}}{V_y}\,.
\label{eqn:tautr}
\end{equation}
If the patch is located on the spherical surface, as depicted in Fig. \ref{fig:Fig7}(c), the interaction time is a function of the rotational velocity $\Omega_x$,
\begin{equation}
\tau_\text{rot} = \frac{2\alpha}{\Omega_x}\,,
\label{eqn:taurot}
\end{equation}
where $2\alpha$ is the angular displacement that corresponds to a linear displacement of $2R_\text{ZOI}$, and $\sin(\alpha) = R_\text{ZOI}/(a + \kappa^{-1})$.
The ``rotational'' interaction time defined by Eq. (\ref{eqn:taurot}) is equivalent to a translational interaction time for small angular displacements ($\sin(\alpha) \approx \alpha$) defined in terms of an effective translational velocity, $V_y^r \equiv a \Omega_x$,
\begin{equation}
\tau_\text{rot} = \frac{2R_\text{ZOI}}{(a + \kappa^{-1})\Omega_x}\,{\approx}\,\frac{2R_\text{ZOI}}{a \Omega_x} = \frac{2R_\text{ZOI}}{V_y^r}\, .
\end{equation}

For particle sizes and Debye lengths frequently chosen in experimental studies \cite{DuffadarDavisSantore:2009, SantoreRolling:2010}, the angular displacements are indeed small. For example, the approximation $\sin(\alpha) \approx \alpha$ presents an error of less than 1\% for particle sizes 2$a$ = 500 nm - 1 $\mu$m and Debye lengths $\kappa^{-1}$ = 2 - 5 nm.
When the particle is not in contact with the collector, it is important to note that $V_y^r < V_y$, as the particle rotates more slowly than it translates \cite{CoxBrennerI:1967, CoxBrennerII:1967}.
The equality $V_y^r = V_y$ holds, however, when the particle rolls without slipping \cite{CoxBrennerI:1967}, as shown in Figs. \ref{fig:Fig4}(c)-(d).

Computations of patch interaction times are most relevant for particle trajectories at surface loadings for which the adhesion probability is small, \textit{i.e.} near the adhesion threshold.
The secondary minimum in the energy-distance profile is chosen as the appropriate separation distance
for the computation of the velocities required in Eqs. (\ref{eqn:tautr})-(\ref{eqn:taurot})
and is computed using the GSI$_\text{UA}$ technique \cite{BenderskyDavis:2011}. The rotational and translational velocities are obtained from the mobility matrix formulation of the hydrodynamics \cite{DuffadarDavis:2007, DuffadarDavisFriction:2008, DuffadarDavisSantore:2009} and the hydrodynamic functions within such matrix are computed with the functional forms given by Duffadar and Davis \cite{DuffadarDavis:2007} at a separation distance $h = D_\text{sec.\,min.}$.

\begin{figure}
\centering
\includegraphics{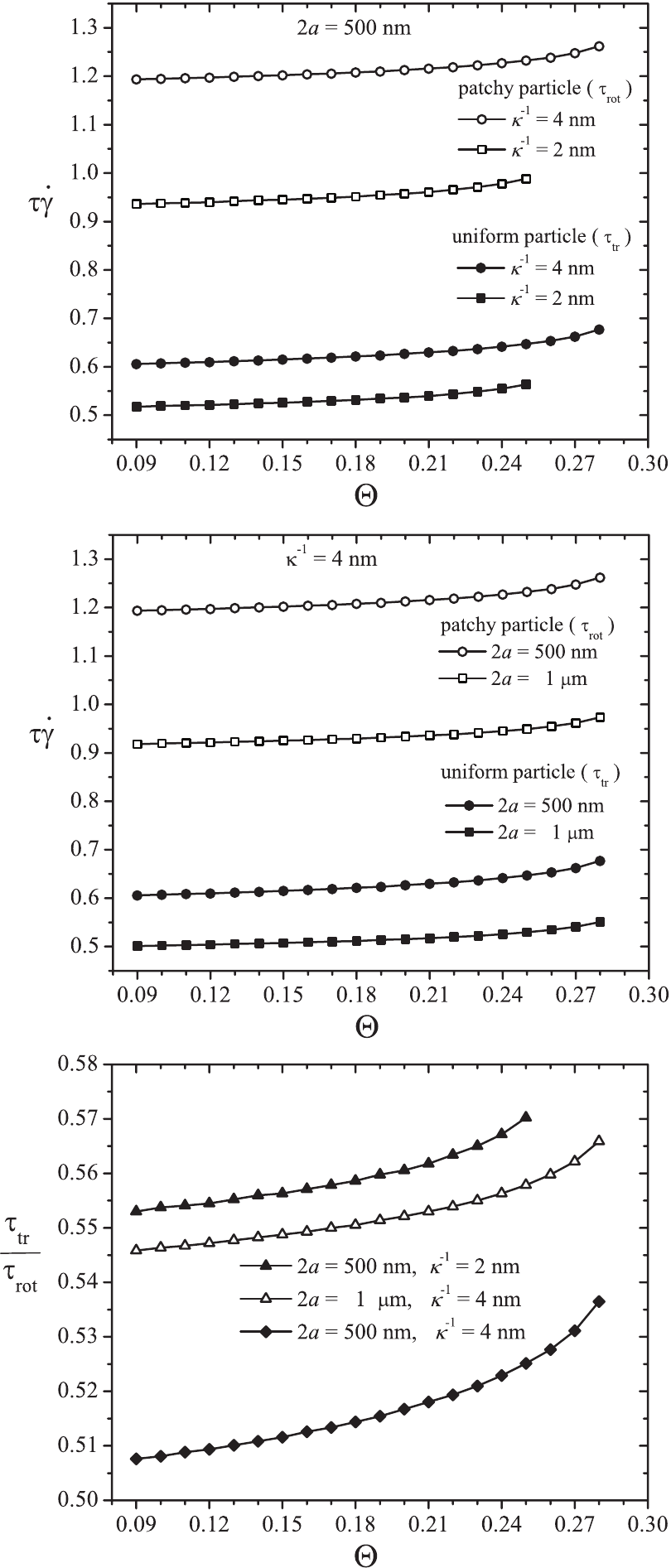} 
\caption{Dimensionless patch interaction times vs. surface loading, for patchy and uniform spheres of varying particle sizes and at different Debye lengths.
(a) Patch interaction times of patchy and uniform particles at Debye lengths $\kappa^{-1} = 2, 4$ nm. The particle size is fixed at $2a$ = 500 nm diameter.
(b) Patch interaction times of patchy and uniform particles, of sizes $2a = 500$ nm and $2a = 1 \mu$m. The Debye length is fixed at $\kappa^{-1} = 4$ nm.
(c) Ratios of translational to rotational patch interaction times for varying particle sizes and Debye lengths.}
\label{fig:Fig8}
\end{figure}

In Fig. \ref{fig:Fig8}, patch interaction times are presented as a function of $\Theta$ for patchy and uniform particles of varying size and at Debye lengths $\kappa^{-1} = 2, 4$ nm. In all cases, the patch interaction time slightly increases with $\Theta$ because stronger attractive interactions decrease the particle-collector separation, which reduces the particle velocities. The values of $\Theta$ considered in patch interaction time calculations are those for which a secondary minimum in the energy-distance profile exists.

For a fixed particle size of 2$a$ = 500 nm, patch interaction times as a function of $\Theta$, for both particle types (patchy, uniform), and two Debye lengths $\kappa^{-1} = 2, 4$ nm are presented in Fig. \ref{fig:Fig8}(a). At a constant Debye length of $\kappa^{-1} = 4$ nm, the interaction times of the patchy particles are significantly larger than those of the uniform spheres (in average, for the presented range of $\Theta$, $\tau_\text{rot} \simeq 1.75\,\tau_\text{tr}$). The same trend is observed for the lower Debye length $\kappa^{-1} = 2$ nm.
The larger patch interaction time of the patchy particle limits heterogeneous (attractive) interactions, because patchy particles interact with each uniform collector element for longer periods of time.
Uniform particles, however, interact with each patch on the collector for shorter periods of time, allowing for an increased number of interactions with multiple patches within the same time period.
In agreement with all the results presented within this work, the larger patch interaction time for patches on the particle translates to a smaller number of trajectory fluctuations (number of local extrema) and lower collection probabilities. It is thus also suggested that adhesion thresholds would be higher for patchy particles.

For either placement of patches, on the particle or on the collector, it is also shown in Fig. \ref{fig:Fig8}(a) that, for a fixed particle size, patch interaction times are larger for the larger Debye length.
At a constant particle size, a change in the Debye length has two competing effects. An increase in the Debye length corresponds to an increase of $R_\text{ZOI} \equiv 2\sqrt{\kappa^{-1} a}$, suggesting the interaction time should increase at larger values of $\kappa^{-1}$. At the same time, however, a larger Debye length moves the secondary minimum farther away from the collector, which increases the particle velocity.  
At a constant particle size of 2$a = 500$ nm, the increase in the size of the ZOI overcomes the increase in the particle velocity, such that the patch interaction times increase with increasing Debye length.

The dependence of the patch interaction time on the surface loading $\Theta$ at a constant Debye length of $\kappa^{-1} = 4$ nm is shown in Fig. \ref{fig:Fig8}(b) for patchy and uniform particles of sizes $2a = 500$ nm and 1 $\mu$m.
For both particle sizes, the patch interaction times of the patchy particle are larger than those of the uniform particles because the rotational velocity is significantly smaller than the translational velocity.  
The effect of particle size on the patch interaction time, at a constant Debye length, is dual, just as the effect of the Debye length at a constant particle size, previously described.
At a fixed Debye length, the larger particle has a larger ZOI, thus, it could be expected that the larger particles will exhibit the larger interaction times. It is the smaller particle, however, for which the interaction times are larger.
The smaller ZOI of the smaller particle defines more localized, attractive interactions that reduce the flowing particles' velocities, to yield larger patch interaction times.

The ratio of the translational and rotational patch interaction times for the particle sizes and Debye lengths presented in Figs. \ref{fig:Fig8}(a)-(b) are shown as a function of $\Theta$ in Fig. \ref{fig:Fig8}(c). It is readily noted that
\begin{equation}
\frac{\tau_\text{tr}}{\tau_\text{rot}} = \frac{2\,R_\text{ZOI}/ V_y}{2\,R_\text{ZOI}/ a\,\Omega_x} = \frac{a\,\Omega_x}{V_y}\;.
\label{eqn:ratio}
\end{equation}
The velocity-, or equivalently, patch interaction time-ratios given by Eq. (\ref{eqn:ratio}) and shown in Fig. \ref{fig:Fig8}(c) are smaller than 1, and thus indicate that the particle translates faster than it rotates, in agreement with the results presented in Figs. \ref{fig:Fig4}(c)-(d) for trajectory segments in which the particle does not contact the collector. It is shown in the Supporting Information that the velocity ratio $a\,\Omega_x/V_y$ increases as the separation $(D/a) \rightarrow 0$, such that smaller separations (due to stronger attractions at larger values of $\Theta$) lead to higher ratios of the interaction times, as shown in Fig. \ref{fig:Fig8}(c).

\subsection{Adhesion regime diagrams}

Interesting particle motions can be distinguished on the basis of system parameters and then presented in adhesion regime diagrams \cite{KornSchwarz:2008, DuffadarDavisFriction:2008}. Duffadar and Davis \cite{DuffadarDavisFriction:2008} presented adhesion regime diagrams as functions of the Debye length and surface loading $\Theta$ in order to distinguish the `arrest', `skipping and rolling', and `no arrest' behaviors of micro-particles in shear flow above a patchy collector.

\begin{figure}
\centering
\includegraphics{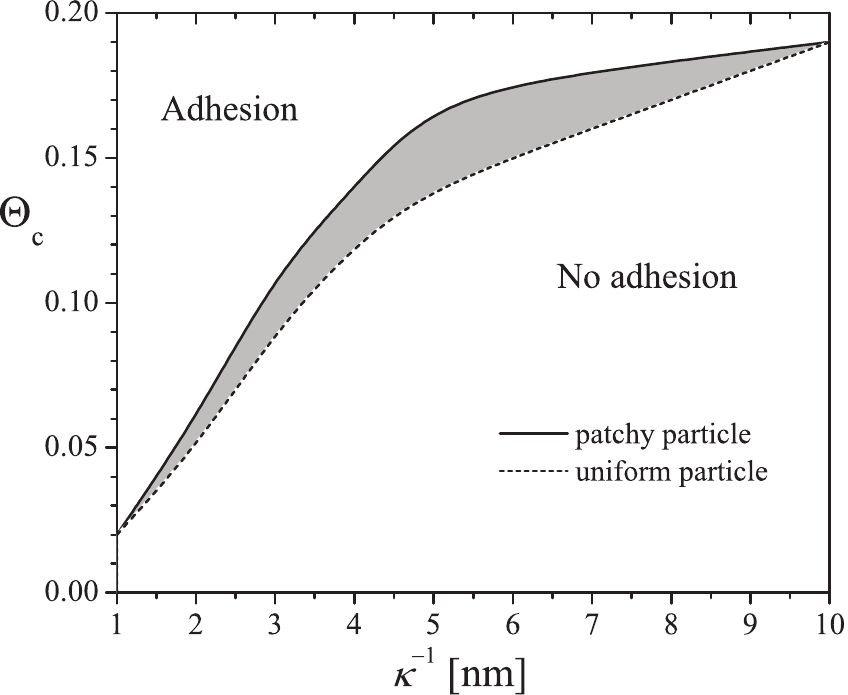}
\caption{Adhesion regime diagram, presented as
the dependence of the adhesion thresholds $\Theta_c$ estimated from collection probability curves on the Debye length $\kappa^{-1}$, for $2a = 500$ nm patchy and uniform particles.}
\label{fig:Fig9}
\end{figure}

Similarly, an adhesion regime diagram is presented in Fig. \ref{fig:Fig9}, on the basis of collection probabilities, such as those computed in Sec. \ref{sec:CollEff}, plotted as a function of $\kappa^{-1}$ and $\Theta$.
The diagram presents adhesion/no adhesion regimes for $2a = 500$ nm patchy and uniform spheres. As expected from the results presented in previous sections, adhesion thresholds estimated from collection probability curves are consistently higher for patchy particles than for uniform ones, such that the adhesion regime of the patchy sphere is smaller than that of the uniform sphere.

The differences in thresholds are more noticeable at relatively low Debye lengths. At a high Debye length of $\kappa^{-1} = 10$ nm, computations within this work revealed an equal threshold of $\Theta_c = 0.19$ for both particle types. Due to a larger ZOI, the heterogeneity is smeared over a larger (planar or spherical) area and the interactions assume a more mean-field like character, irrespective of their specific location on the collector or sphere.

It is interesting to note that each of the `adhesion' and `no adhesion' regimes can be subdivided to account for rolling/skipping motions and thus distinguish between 4 typical dynamic behaviors, given by the adhesion/no adhesion and surface contact/no surface contact possible combinations.
Particles of $2a = 500$ nm diameter, with a Debye length $\kappa^{-1} = 5$ nm (upper lines in Fig. \ref{fig:Fig4}(b) are examples of particles that do not contact the collector and do not adhere on the surface (patchy sphere) and that do not contact the collector but do adhere on the surface (uniform sphere). Alternatively, at a low Debye length $\kappa^{-1} = 1$ nm, small patchy and uniform spheres contact the collector and adhere on it (lower lines in Fig. \ref{fig:Fig4}(b)), while large particles contact the collector but without adhering (lower lines in Fig. \ref{fig:Fig4}(a)).

\section{Conclusion}
\label{sec:concl}

Particle-collector systems with one heterogeneous surface, which is either the spherical particle or the planar collector, are characterized in detail for different system geometries and
a range of particle sizes, Debye lengths and electrostatic potentials.
DLVO interactions are computed by implementing the GSI technique, and incorporated in a mobility matrix formulation of the dynamics problem that yields particle trajectories as the particle translates in shear flow above a flat collector.

The (particle)-(patchy collector) system, thoroughly studied in previous work, does not require spherical surfaces to be discretized into areal surface elements. To model interactions of systems that include patchy particles, however, the inclusion of such a discretization scheme in the computational model is, in fact, essential.
The use of the recursive zonal EQual area Sphere Partitioning (EQSP) algorithm for the discretization of spherical particles into small, equal-area elements is validated through energy computations of the uniform particle-collector system. 
DLVO interactions for (patchy particle)-(collector) systems are thus modeled by incorporating EQSP-generated spherical surface elements within the GSI technique. The system's dynamic behavior is also obtained in this case from mobility matrix computations.

Differences in the adhesive and dynamic behaviors of particle-collector patchy systems, in which only \emph{one} surface is patterned with nanoscale features, are quantified by computations of collection probabilities and of average numbers of trajectory local extrema. Patch interaction times in each case are also defined.

The lessened tendency of the patchy particle to adhere on a uniform collector, with respect to that of the uniform particle adhering on a patchy collector, is attributed to larger patch interaction times for the patchy particle. A larger interaction time precludes multiple interactions with many heterogeneous surface elements in a given time period, which translates into fewer attractive interactions.
Moreover, larger interaction times reduce the amount of spatial fluctuations exhibited by patchy particles interacting with uniform collectors, and lead to a smaller number of local extrema in the trajectory, as shown in the results presented in this work.
Spatial variations in the trajectory, in turn, correlate with the extent of interactions with heterogeneous surface elements and ultimately provide insight into the adhesive character of the system for a given set of parameters. Adhesion thresholds estimated from collection probability curves of patchy particles are indeed larger than those of
uniform particles adhering on patchy collectors.

In summary, a new computational approach is introduced in this work to compute particle trajectories, along with statistical measures that characterize those trajectories, of patchy particles flowing above uniform planar collectors.
The EQSP discretization scheme of spherical surfaces is incorporated in GSI computations of DLVO interactions and in the mobility matrix formulation that yields particle trajectories. 
The use of this newly developed simulation technique can be extended, for instance, to model particle-collector systems of \emph{two} heterogeneous surfaces, and of surfaces covered with many types of heterogeneity, such as patches, pillars, and spring-like structures that resemble polymer brushes or cellular receptors, which will be pursued in future work.

\section{Acknowledgements}
We gratefully acknowledge support from UMass MRSEC on Polymers, NSF-0820506 and NSF-1264855.

\bibliographystyle{plain}

\end{document}